\documentclass[lettersize,journal]{IEEEtran}
\usepackage{amsmath,amsfonts}
\usepackage{algorithmic}
\usepackage{algorithm}
\usepackage{amsthm}
\usepackage{amssymb}
\usepackage{booktabs}
\usepackage{xcolor}

\theoremstyle{definition}

\theoremstyle{theorem}

\theoremstyle{lemma}

\theoremstyle{Assumption}

\usepackage{array}
\usepackage[caption=false,font=normalsize,labelfont=sf,textfont=sf]{subfig}
\usepackage{textcomp}
\usepackage{stfloats}
\usepackage{url}
\usepackage{verbatim}
\usepackage{graphicx}
\usepackage{cite}
\usepackage{color}
\usepackage{soul}

\begin{document}
\title{Stacked Intelligent Metasurface-Based MultiModal Feature Fusion for Collaborative Inference in Edge Networks}
\author{Yushi Wang, Zhengxin Yu,~\IEEEmembership{Member,~IEEE}, Guhan Zheng,~\IEEEmembership{Member,~IEEE}, Haris Pervaiz,~\IEEEmembership{Senior Member,~IEEE},\\Zheyi Chen,~\IEEEmembership{Member,~IEEE}, and Shunqing Zhang,~\IEEEmembership{Senior Member,~IEEE}    


\thanks{Yushi Wang and Zhengxin Yu are with the School of Computing and Communications, Lancaster University, LA1 4YW, United Kingdom. (e-mail: y.wang216@lancaster.ac.uk; z.yu8@lancaster.ac.uk).}
\thanks{Guhan Zheng and Shunqing Zhang are with the School of Communication and Information Engineering, Shanghai University, 200444, China. (e-mail: gzheng@shu.edu.cn; shunqing@shu.edu.cn).}
\thanks{Haris Pervaiz is with the School of Computer Science and Electronic Engineering, University of Essex, CO4 3SQ, UK. (e-mail: haris.pervaiz@essex.ac.uk).}
\thanks{Zheyi Chen is with the College of Computer and Data Science, Fuzhou University, Fuzhou 350002, China. (e-mail: z.chen@fzu.edu.cn)}
}

\maketitle

\begin{abstract}
In navigating the challenges of computational and communication overhead within the 6G-era Collaborative Inference (CI), this paper explores the feasibility of employing Stacked Intelligent Metasurface (SIM) to perform CI, termed SIM-CI. 
The SIM functions as a physical computing unit, receiving multi-user Over-The-Air (OTA) aggregated features within the same spectrum. It thus executes the necessary spatial-frequency combining and inference in a single physical wave propagation. 
SIM-CI also incorporates a tri-heterogeneity robustness adapter to address the network heterogeneity, such as multi-modal data and user inference models, as well as a quality-aware contrastive approach that align features from different modalities and extraction levels into a unified feature space. 
Considering the heterogeneous and dynamic channels, we further integrate Channel State Information (CSI) as an additional input into the adapter design, enabling users' early perception of potential channel states. Moreover, to maximally mitigate the impact of channel variations, we design a new hybrid beamforming network that combines the robustness of model-driven pre-equalization with the adaptability of data-driven learning. 
Experimental results demonstrate the effectiveness of our SIM-CI compared to digital-domain inference, achieving 98.6\% accuracy to conversional digital-domain CI approaches under various heterogeneous conditions, while reducing
computational energy by 25\% and requiring 75\% fewer transmission time slots.
\end{abstract}


\section{Introduction}

Sixth-generation (6G) wireless networks aim to support ubiquitous intelligence \cite{liu2024near}, where massive numbers of distributed mobile and edge devices, such as autonomous vehicles, unmanned aerial vehicles, and Internet-of-Things devices, require real-time reasoning and decision-making capabilities with the support of edge intelligence \cite{zheng2024mobility, zheng2024semantic}. By splitting neural network models, part of an inference task can be executed locally, while the remaining computation is offloaded to the network edge, balancing device capabilities and task demands. Recently, multi-user Collaborative Inference (CI) has further emerged as a promising paradigm \cite{yang2022semantic,dai2022nonlinear}. For the same inference task, edge infrastructure can aggregate multimodal information collected by heterogeneous users, providing a broader sensing scope and more robust collaborative decision-making while reducing redundant computation. However, in mobile edge networks, distributed users continuously generate and upload heterogeneous intermediate features, which consumes substantial spectrum resources \cite{yang2023energy}. Digital-domain feature fusion and joint inference at the edge also introduce considerable computation latency and energy consumption \cite{zhu2021over}.

Breakthroughs in the field of Programmable Metasurfaces (PMs), particularly technologies represented by the Stacked Intelligent Metasurface (SIM) \cite{an2023stacked}, offer an innovative route to address the physical bottlenecks described above. SIM directly executes computational tasks during the physical propagation of the signal by programming and controlling electromagnetic waves  \cite{huang2024stacked,huang2025stacked}. This analog computing paradigm has the potential to reduce computation latency and energy consumption introduced by edge digital-domain processing in CI.


\subsection{Challenges}


SIM can provide low latency and low energy for CI, but it also implies that edge networks lose the ability to align, filter, and weight heterogeneous information in the digital domain. This is because SIM replaces the flexible digital processing modules in edge servers with analog superposition at the physical layer. This challenge can become more pronounced in mobile edge networks, where user conditions and available resources may vary over time. Furthermore, given the widespread presence of data heterogeneity, model heterogeneity, and system heterogeneity in the real world \cite{kairouz2021advances, Zheng2025}, the physical computing process based on SIM faces additional obstacles and may render CI results uncontrollable.

As a result of data heterogeneity, different users may require similar inference outputs even though their data distributions and content modalities differ. For example, image features produced by onboard cameras and text features collected by roadside units reside in different feature spaces. Directly superimposing these heterogeneous features during signal propagation may cause feature aliasing rather than effective information fusion.


Furthermore, model heterogeneity leads to inconsistent feature extraction depths across users with different computing capabilities. Some users may provide shallow texture features while others provide deep abstract concepts. Directly computing these features of varying abstraction in SIM can bias the final inference result. In mobile networks, changes in available computation resources may also cause the feature extraction depth of a user to vary between uploads through early-exit mechanisms \cite{teerapittayanon2016branchynet}.


%
Channel heterogeneity and dynamic variations also pose severe challenges for transmitting heterogeneous features in SIM. Different users experience different wireless channel conditions that fluctuate over time. Thus, the channel ceases to be a passive transmission medium and becomes an uncontrolled analog computation layer that imposes random multiplicative distortion on the transmitted feature vectors. Specifically, channel-induced amplitude fading and phase rotation can destroy the geometric structure of feature vectors, distorting their relative positions and relationships in feature space and thereby causing large offsets in CI results. Especially when SIM serves as the edge node for CI, the lack of intelligent filtering of low-quality features that edge servers normally provide means that corrupted features will be included in the fusion process, undermining the integrity and reliability of inference results. 


Existing studies have investigated these data, model and system heterogeneity issues mainly in conventional digital-domain CI and edge computing systems through cross-modal alignment \cite{chen2020uniter,cheng2022cross,radford2021learning}, adaptive inference \cite{bolukbasi2017adaptive,zhao2022edgeadaptor,peng2024distree}, and predictive or robust beamforming \cite{liu2020radar,shi2023robust,yuan2020transfer}. These approaches mainly address individual sources of heterogeneity and rely on flexible digital processing. Such flexibility is limited when multi-user features are directly superimposed and processed through SIM, making the joint handling of data, model, and channel heterogeneity challenging.

\subsection{Motivation and Contributions}


To bridge the gap identified above, we propose a SIM-based CI framework, \emph{i.e.}, SIM-CI. It deploys the SIM at the network edge and embeds part of the inference into the physical domain of electromagnetic wave propagation, thereby enabling efficient migration of inference tasks from the digital domain to the physical domain. In SIM-CI, a tri-heterogeneity robustness adapter is designed to dynamically coordinate the mapping between heterogeneous data and heterogeneous models to improve inference robustness. Furthermore, to cope with the heterogeneous and time-varying characteristics of wireless channels, SIM-CI integrates a dual-robustness mechanism that combines implicit sensing and explicit compensation to further enforce inference accuracy in the physical domain. The main contributions of this paper are summarized as follows:

\begin{itemize}

\item We propose SIM-CI, a novel physical-domain collaborative inference architecture that seamlessly integrates transmission and computation. SIM-CI preprocesses user feature vectors through a designed tri-heterogeneity robustness adapter and beamforming networks at the user side, allowing multiple users to transmit features in parallel over the same frequency band. By leveraging Over-the-Air (OTA) computation, these features are aggregated in the analog domain and directly processed by the SIM to perform inference on the superimposed waveform, realizing an integrated transmission-as-computation paradigm. Theoretically, this framework enables inference at the speed of light without incurring digital computation energy consumption, fundamentally reducing the excessive communication and computation overheads in conventional digital collaborative inference systems.


\item We design a tri-heterogeneity robustness adapter to address data and model heterogeneity in physical-domain computation. Built on Graph Neural Networks (GNN), the adapter constructs a graph similarity matrix integrating intra-modal and cross-modal relationships to unify heterogeneous graph structures and reduce semantic distances among multimodal features in the latent space. Moreover, we develop a quality-aware contrastive learning mechanism that quantifies features extracted from heterogeneous models into quality indicators and incorporates them into intra-modal and cross-modal contrastive losses. By assigning greater weight to high-quality features, the proposed design aligns features from different modalities and abstraction levels into a unified feature space, enhancing the accuracy and robustness of physical-domain superposition computation.


\item We propose a dual-robustness mechanism that combines implicit channel sensing and explicit channel compensation to reduce the side effect of channel distortion on SIM-based CI. In the implicit sensing stage, predicted Channel State Information (CSI) is provided as an additional input to the adapter, enabling feature alignment to anticipate channel variations and build intrinsic anti-interference capability in the feature space. To further address residual distortion, we design a novel hybrid beamforming network for explicit compensation with two branches: 1) a model-driven branch performs phase pre-alignment and amplitude pre-equalization for proactive channel compensation and accurate signal superposition; 2) a data-driven branch utilizes a self-attention mechanism to capture complex inter-subcarrier dependencies and learn an adaptive end-to-end mapping function. The proposed dual mechanism encodes channel-aware features produced by the adapter into highly robust transmission waveforms, achieving efficient OTA aggregation while significantly reducing the negative impact of channel heterogeneity.

\end{itemize}

The rest of the paper is structured as follows: Section \ref{relatedwork} reviews the related work. Section \ref{Sec:sysmodel} presents the system model. In Section \ref{Sec:framework}, we describe the SIM-CI framework in details. The performance evaluation of SIM-CI is provided
in Section \ref{Sec:evaluation}. Finally, Section \ref{Sec:conclusion} concludes this paper.

\section{Related Works}
\label{relatedwork}

Most existing approaches for handling heterogeneous features, computing capabilities, and wireless conditions still rely on conventional digital processing. UNITER \cite{chen2020uniter} learns joint image-text representations by modeling fine-grained correspondences between visual regions and text tokens, while cross-modal graph matching \cite{cheng2022cross} instead captures the structural relations between the two modalities. VSE++ \cite{faghri2017vse++} and DeViSE \cite{frome2013devise} take a simpler route, mapping images and text directly into a shared semantic space. Fragment-level self-attention \cite{wu2019learning} and large-scale contrastive learning \cite{radford2021learning} push this idea further by modeling context and semantic correspondence explicitly. These methods align heterogeneous modalities well, but they generally assume the aligned features share a similar level of abstraction. That assumption breaks down once users' computing budgets start to differ.

A separate line of work considers the heterogeneous computing capabilities of edge devices. Adaptive neural networks \cite{bolukbasi2017adaptive} adjust computation through early exits or network selection. EdgeAdaptor \cite{zhao2022edgeadaptor} goes a step further, handling online model configuration, model selection, and resource provisioning together for edge DNN inference, while DistrEE \cite{peng2024distree} distributes early-exit inference across multiple edge devices. These studies mainly focus on adapting local computation to device
capabilities, while the fusion of features extracted at different
depths at a physical receiver such as SIM remains less explored.

Even if model heterogeneity is handled in the digital domain, dynamic wireless channels can still throw aligned features out of balance before they reach the receiver. Channel variation is a well-studied problem on its own. Predictive beamforming adapts transmission to fast-changing vehicular channels \cite{liu2020radar}, and robust WMMSE precoding accounts for channel uncertainty under imperfect CSI \cite{shi2023robust}. Other work attacks the signal directly, through power control \cite{lee2018deep} or channel equalization \cite{lv2023deep,restuccia2020deep}. Transfer and meta learning have also been used to help beamformers adapt as the environment shifts \cite{yuan2020transfer}, and some end-to-end schemes go further still, jointly optimizing source and channel coding \cite{bourtsoulatze2019deep} or folding beamforming directly into the semantic coding pipeline \cite{wu2024deep}.

These approaches improve transmission robustness under dynamic wireless conditions, but the channel plays a different role in SIM-based CI. In conventional systems, channel adaptation mainly protects the information being transmitted. In SIM-based CI, channel distortion directly changes the superposition of multi-user features and therefore the input to the physical computing process. As a result, channel robustness must be considered together with feature alignment rather than only during signal transmission.

SIM has also been investigated for multiuser signal processing and semantic communication. An et al. \cite{an2025stacked} jointly optimized wave-domain beamforming and transmit power allocation for multiuser downlink communications. Papazafeiropoulos et al. \cite{papazafeiropoulos2025channel} developed an MMSE channel estimation protocol for SIM-assisted systems under Rician fading. These studies address communication performance and channel acquisition, rather than the fusion of heterogeneous semantic features. For task-oriented inference, Huang et al. \cite{huang2024stacked} employed SIM to perform source and semantic encoding for image recognition. Their subsequent work \cite{huang2025stacked} combined visual and textual semantic information for scene reconstruction. Li et al. \cite{li2026distributed} further developed distributed electromagnetic neural networks, where multiple UAV-mounted SIMs collaboratively encode image semantics for over-the-air aggregation and recognition. Although these studies demonstrate multimodal transmission and distributed wave-domain processing, they do not jointly address the alignment of features from different modalities and local model depths together with channel compensation for multiuser inference. The proposed SIM-CI framework is designed to address this joint design problem.

\section{System Model} \label{Sec:sysmodel}

\begin{figure*}[t]
    \centering
        \centering
        \includegraphics[width=\textwidth]{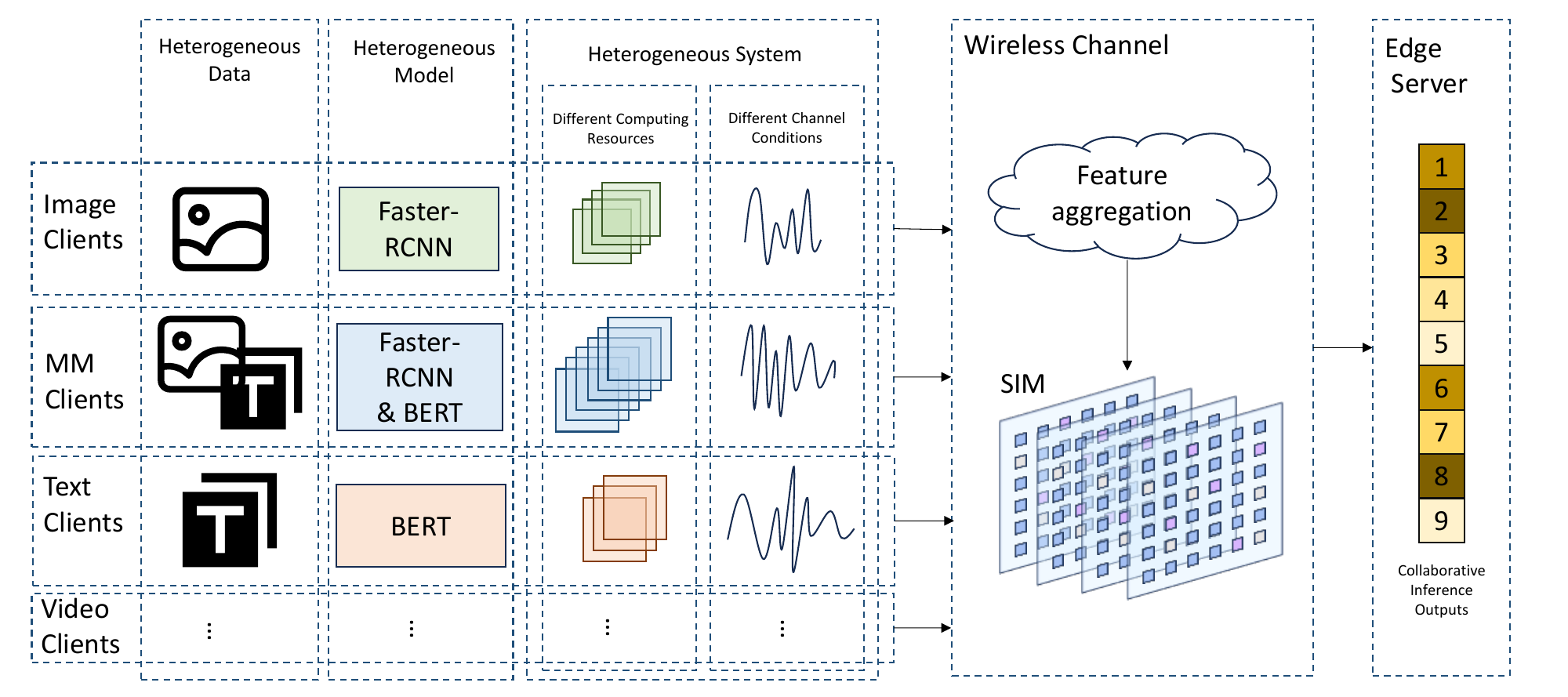}
    \caption{The system model for multi-user CI, illustrating the challenges of data, model, and system heterogeneity under the proposed SIM-CI physical computing paradigm.}
    \label{fig:0}
    \centering
\end{figure*}

We consider a multi-user CI system comprising $K$ single-antenna
users and an edge server equipped with an $N_r$-antenna uniform
planar array (UPA). The users provide heterogeneous observations
of a common scene and collaboratively infer the object categories
present in that scene. The system employs OFDM with $N_c$
subcarriers. The CI workflow consists of the following phases:

\subsection{Data Acquisition and Feature Extraction}
Each user \( k \in \{1, \ldots, K\} \) locally acquires heterogeneous raw data, denoted as \( \mathcal{D}_k \), which may include visual images \( \mathbf{I}_k \) or textual descriptions \( \mathbf{T}_k \), etc. User \( k \) then employs its heterogeneous model to perform part of the inference process and abstract the raw data into a structured set of features. For instance, image and text inputs are processed by the local
encoders $E_k^{\mathrm{img}}(\cdot)$ and
$E_k^{\mathrm{txt}}(\cdot)$, respectively. Their outputs are
mapped to a common dimension $d$ through learnable linear
projections, yielding visual tokens
$\{\mathbf{r}_i^k\}_{i=1}^{n_r^k}$ and textual tokens
$\{\mathbf{c}_j^k\}_{j=1}^{n_c^k}$, where
$\mathbf{r}_i^k,\mathbf{c}_j^k\in\mathbb{R}^{d}$.

\subsection{Wireless Communication Model}
The extracted features are channel-encoded and modulated into complex symbol vectors \( \mathbf{s}_k \) for uplink transmission. The considered multi-user OFDM system is with subcarrier spacing \( \Delta f \) and sampling period \( T_s = 1/(N_c \Delta f) \).

The wireless channel from user \( k \) to the server is modeled as a frequency-selective channel with \( L_k \) distinct propagation paths in a rich scattering environment. The frequency-domain channel vector on subcarrier \( n \) is given by \cite{gao2022data}:
\begin{equation}
\label{eq:channel_model}
\mathbf{h}_k[n] = \frac{1}{\sqrt{L_k}} \sum_{\ell=1}^{L_k} \alpha_{k,\ell} \cdot \mathbf{a}_r(\theta_{k,\ell}, \phi_{k,\ell}) \cdot \exp\left(-j \frac{2\pi n \tau_{k,\ell}}{N_c T_s}\right) \in \mathbb{C}^{N_r},
\end{equation}
where \( \alpha_{k,\ell} \sim \mathcal{CN}(0, \sigma_\alpha^2) \) is the complex gain of the \( \ell \)-th path, \( \tau_{k,\ell} \) is the path delay, and \( (\theta_{k,\ell}, \phi_{k,\ell}) \) is the angle of arrival.

The UPA response vector \( \mathbf{a}_r(\theta,\phi) \in \mathbb{C}^{N_r} \) is formulated as:
\begin{equation}
\label{eq:array_response}
\mathbf{a}_r(\theta,\phi) = \mathbf{a}_x(\theta,\phi) \otimes \mathbf{a}_y(\theta,\phi),
\end{equation}
where \( \otimes \) denotes the Kronecker product, and the individual array responses for the \( N_x \times N_y \) UPA (with \( N_r = N_x N_y \)) are:
\begin{align}
\label{eq:array_x}
\mathbf{a}_x(\theta,\phi) &= \left[1, e^{j\pi \sin\phi \cos\theta}, \ldots, e^{j\pi (N_x-1)\sin\phi \cos\theta}\right]^T, \\
\label{eq:array_y}
\mathbf{a}_y(\theta,\phi) &= \left[1, e^{j\pi \sin\phi \sin\theta}, \ldots, e^{j\pi (N_y-1)\sin\phi \sin\theta}\right]^T.
\end{align}

\subsection{Conventional Edge Processing}
The server receives the signals from all \( K \) users. It first executes channel decoding and demodulation to recover estimates of the transmitted features, denoted as \( \{\hat{\mathbf{r}}_i^k\} \) and \( \{\hat{\mathbf{c}}_j^k\} \). It then executes a CI model \( g(\cdot) \) that fuses these heterogeneous features to produce the final inference result as:
\begin{equation}
\label{eq:inference_model}
\mathbf{y} = g\left( \{\hat{\mathbf{r}}_i^k\}_{i=1}^{n_r^k}, \{\hat{\mathbf{c}}_j^k\}_{j=1}^{n_c^k} \right)_{k=1}^K.
\end{equation}

\section{The Proposed SIM-CI Framework }\label{Sec:framework}
In this section, we present the model of our proposed SIM-CI. We then detail the core designs for mitigating various network heterogeneities.

\subsection{SIM-CI design}
The SIM consists of $L_{\text{rx}}$ layers, where fixed physical propagation between layers acts as a linear transformation and trainable meta-atoms apply controllable complex-valued modulation  \cite{lin2018all,liu2022programmable}. By optimizing the physical states of the meta-atoms, the SIM is trained end-to-end to process aggregated multi-user features for collaborative inference.

\subsubsection{Data Acquisition and Feature Extraction}

Similar to general CI, each user locally acquires heterogeneous raw data. User $k$ then employs its heterogeneous model to perform part of the inference process and abstract the raw data into a structured set of features. Nevertheless, without the intelligent computing capabilities of the edge, the SIM-CI struggles to process heterogeneous features from multi-modal row data and different model depth levels. We hence designed a novel robust adapter based on GNN. This adapter takes as input the user's multi-modal token sequences ($\{\mathbf{r}_i^k\}$, $\{\mathbf{c}_j^k\}$), the feature quality indicators ($\text{depth}_{\text{img}}^k$, $\text{depth}_{\text{txt}}^k$), and the user's estimated channel vectors $\{\mathbf{\hat{h}}_k[n]\}_{n=1}^{N_c}$, where $N_c$ is the number of OFDM subcarriers and $\mathbf{\hat{h}}_k[n] \in \mathbb{C}^{N_r}$ with $N_r$ being the number of receive antennas. Here, the indicators $\text{depth}_{\text{img}}^k \in [1, L_{\text{max}}]$ (where $L_{\text{max}}$ is the encoder's maximum depth) quantify the extraction depth of the input tokens. The role of the adapter is to project these disparate inputs into a single, unified feature vector $\mathbf{g}_k \in \mathbb{R}^{D_{\text{sem}}}$, where $D_{\text{sem}}$ is the semantic feature dimension. This vector $\mathbf{g}_k$ represents an aligned, channel-aware semantic representation, independent of the original data modality or feature depth. The aligned feature $\mathbf{g}_k$ is passed to a designed beamforming network tailored for SIM-CI. This network aims to address the channel heterogeneity and map the real-valued feature vector $\mathbf{g}_k$ to a complex-valued symbol vector $\mathbf{s}_k = [s_k[1], \ldots, s_k[N_c]]^T \in \mathbb{C}^{N_c}$ for OFDM transmission. It takes $\mathbf{g}_k$ and the channel estimate $\hat{\mathbf{h}}_k[n]$ as inputs. We consider an imperfect CSI model where the estimated channel $\hat{\mathbf{h}}_k[n]$ is given by $\hat{\mathbf{h}}_k[n] = \mathbf{h}_k[n] + \mathbf{e}_k[n]$. The estimation error $\mathbf{e}_k[n] \sim \mathcal{CN}(\mathbf{0}, \sigma_e^2 \mathbf{I}_{N_r})$ accounts for practical impairments such as pilot contamination and feedback delays. The output $\mathbf{s}_k$ is optimized to jointly encode the semantic information and pre-compensate for channel distortions.

\subsubsection{Wireless Communication Model}
Although SIM has the potential to reduce the computing delay and energy cost, it still faces numerous uplink spectrum resource demands for inference. To further enhance spectrum efficiency during feature transmission, we ingeniously leverage the SIMs' over-the-air (OTA) computing characteristics. This allows users to transmit simultaneously over the same frequency band before CI, achieving OTA aggregation of multi-user features. The signals from $K$ users naturally superimpose in the wireless multiple access channel. The received signal at the server on subcarrier $n$ is:
\begin{equation}
\label{OTA_agg}
\mathbf{y}[n] = \frac{1}{\sqrt{K}} \sum_{k=1}^{K} \mathbf{h}_k[n] \cdot s_k[n] + \mathbf{z}[n] \in \mathbb{C}^{N_r},
\end{equation}
where $\mathbf{z}[n] \sim \mathcal{CN}(\mathbf{0}, \sigma_n^2 \mathbf{I}_{N_r})$ is additive white Gaussian noise (AWGN) with noise power $\sigma_n^2$, and the normalization factor $1/\sqrt{K}$ ensures proper scaling of received power. This physical superposition process is the core of OTA aggregation computation. The wireless channel itself acts as an analog adder, directly fusing features from multiple users in the electromagnetic domain. However, the random amplitude fading and phase rotation from each user's channel $\mathbf{h}_k[n]$ can destroy the geometric structure of the features. If channel variations are large, features from weak channels will be overwhelmed. If phases are misaligned, feature vectors will randomly rotate in the complex plane, leading to aggregation failure. This is precisely why the beamforming network must perform pre-compensation at the transmitter. By using amplitude pre-equalization and phase pre-alignment, the network compensates for channel phase rotation and regularizes
the transmitted amplitudes. This process allows the aggregated features to maintain their correct geometric relationships.

Concatenating the received signals across all subcarriers yields the complete space-frequency signal matrix:
\begin{equation}
\label{OTA_receive}
\mathbf{Y} = [\mathbf{y}[1]^T, \mathbf{y}[2]^T, \ldots, \mathbf{y}[N_c]^T]^T \in \mathbb{C}^{N_c \times N_r},
\end{equation}
where each row corresponds to the $N_r$-dimensional received vector on one subcarrier. This matrix contains all information from $K$ users after feature alignment, beamforming, channel propagation, and physical aggregation, and serves as the input to the receiver SIM.

\subsubsection{Novel SIM processing}

We deploy the SIM at the edge to function as the core of our physical inference engine. It receives the complete OTA-aggregated space-frequency signal matrix, $\mathbf{Y} \in \mathbb{C}^{N_c \times N_r}$, as its direct analog input. The SIM is trained to function as a powerful, high-dimensional linear classifier. It processes the incoming $M$-dimensional ($M = N_c \times N_r$) wavefront and is optimized to map different aggregated features to $C$ distinct output ports, where $C$ is the number of target classes. At these output ports, energy detectors measure the signal intensity, yielding the raw classification logits $\{p_c\}_{c=1}^C$. These logits are then passed to a lightweight digital classification head to produce the final normalized class probabilities, thus completing the entire CI task.

\begin{figure*}[t]
    \centering
        \centering
        \includegraphics[width=1\textwidth]{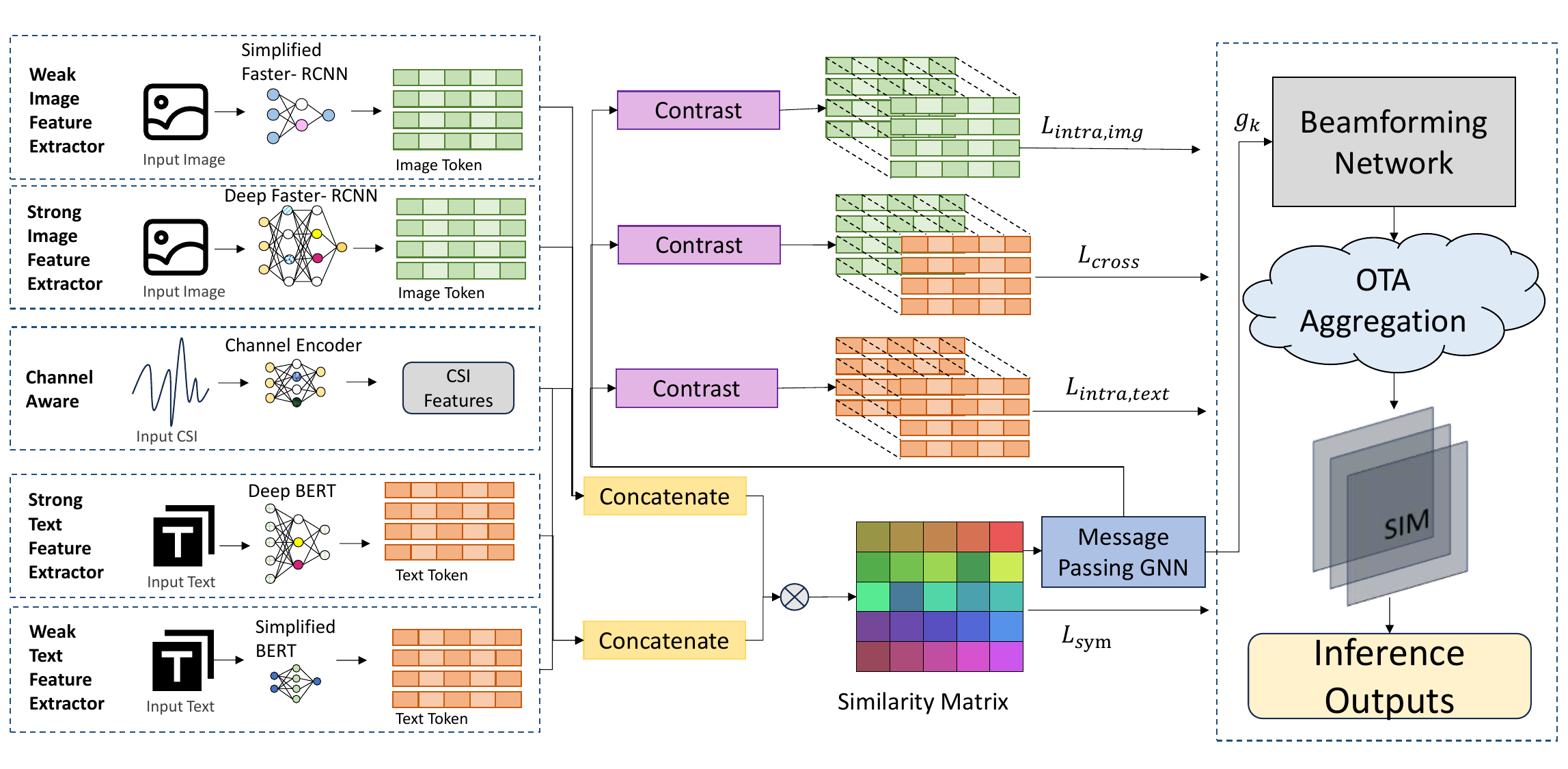}
    \caption{Detailed architecture of the proposed SIM-CI framework, illustrating the tri-heterogeneity robustness adapter, the beamforming network, OTA aggregation, and SIM receiver for final classification.}
    \label{fig:1}
    \centering
\end{figure*}

\subsection{Tri-heterogeneity Robustness Adapter based on GNN}
The proposed adapter is designed for each user to extract a unified feature representation from both heterogeneous data and the outputs of heterogeneous models. It enables robust feature alignment across diverse modalities and abstraction levels. To clarify, we use text and image as input data modalities to illustrate the SIM-CI process. The proposed graph construction is not restricted to two modalities. For a client with multiple available modalities, the similarity matrix is extended with the corresponding intra-modal and pairwise cross-modal blocks, while the pooled representations of all available modalities are concatenated to form the fused feature. The cross-modal contrastive objective is similarly applied to all available modality pairs.

We consider the user $k$ who has raw visual tokens $\{\mathbf{r}_i\}_{i=1}^{n_r}$ and text tokens $\{\mathbf{c}_j\}_{j=1}^{n_c}$. Here, $n_r$ and $n_c$ are the number of visual and text tokens, and each feature vector $\mathbf{r}_i, \mathbf{c}_j$ is in $\mathbb{R}^{d}$.

The core of the adapter is a GNN, which is designed to align these disparate tokens. To illustrate the GNN's core modeling logic for handling data and model heterogeneity, let us first define a generic set of initial node representations $\mathbf{f}^{(0)}$. We define them directly from the raw features:
\begin{equation}
\begin{aligned}
\mathbf{f}_{\text{img},i}^{(0)} &= \mathbf{r}_i \in \mathbb{R}^{d}, \quad i = 1, \ldots, n_r, \\
\mathbf{f}_{\text{txt},j}^{(0)} &= \mathbf{c}_j \in \mathbb{R}^{d}, \quad j = 1, \ldots, n_c.
\end{aligned}
\end{equation}
The adapter operates on a heterogeneous graph of visual and text tokens. To define the graph's structure, we build a similarity matrix $\mathbf{S} \in \mathbb{R}^{(n_r+n_c) \times (n_r+n_c)}$ based on these initial features $\mathbf{f}^{(0)}$ to capture both intra-modal and cross-modal relationships.

For intra-modal similarity, we capture the local structure using a k nearest neighbor (kNN) constrained Radial Basis Function (RBF) kernel. For visual tokens, the similarity is:
\begin{equation}
(\mathbf{S}_{VV})_{ij} = \begin{cases}
\exp\left(-\frac{\|\mathbf{f}_{\text{img},i}^{(0)} - \mathbf{f}_{\text{img},j}^{(0)}\|^2}{2\sigma^2}\right), & \text{if } j \in \mathcal{N}_{knn}(\mathbf{f}_{\text{img},i}^{(0)}), \\
0, & \text{otherwise},
\end{cases}
\end{equation}
where $\mathcal{N}_{knn}(\cdot)$ is the set of $k$ nearest neighbors and $\sigma > 0$ is the RBF kernel bandwidth. The RBF kernel assigns decaying weights to nearby nodes, giving closer features stronger influence. The kNN constraint makes the graph sparse, which reduces computation and preserves the local feature structure. The text token similarity $\mathbf{S}_{TT}$ is computed in the same way:
\begin{equation}
(\mathbf{S}_{TT})_{ij} = \begin{cases}
\exp\left(-\frac{\|\mathbf{f}_{\text{txt},i}^{(0)} - \mathbf{f}_{\text{txt},j}^{(0)}\|^2}{2\sigma^2}\right), & \text{if } j \in \mathcal{N}_{knn}(\mathbf{f}_{\text{txt},i}^{(0)}), \\
0, & \text{otherwise}.
\end{cases}
\end{equation}

For cross-modal similarity, we bridge the two modalities. We first compute a raw affinity matrix $\mathbf{A} \in \mathbb{R}^{n_r \times n_c}$ using inner products:
\begin{equation}
\mathbf{A}_{ij} = (\mathbf{f}_{\text{img},i}^{(0)})^T \mathbf{f}_{\text{txt},j}^{(0)},
\end{equation}
where $\mathbf{A}_{ij}$ measures the alignment between visual token $i$ and text token $j$. To highlight the best matches while keeping the global context, we combine $\text{TopK}$ selection with average pooling. $\text{TopK}(\mathbf{A}, k_{cross})$ selects the top $k_{cross}$ values in each row, and $\text{MeanPool}(\mathbf{A})$ computes row averages. These are refined by MLPs and added:
\begin{equation}
\mathbf{S}_{VT} = \text{MLP}_{VT}(\text{TopK}(\mathbf{A}, k_{cross})) + \text{MeanPool}(\mathbf{A}),
\end{equation}
\begin{equation}
\mathbf{S}_{TV} = \text{MLP}_{TV}(\text{TopK}(\mathbf{A}^T, k_{cross})) + \text{MeanPool}(\mathbf{A}^T).
\end{equation}
The final similarity matrix $\mathbf{S}$ assembles these four blocks:
\begin{equation}
\mathbf{S} = \begin{bmatrix}
\mathbf{S}_{VV} & \mathbf{S}_{VT} \\
\mathbf{S}_{TV} & \mathbf{S}_{TT}
\end{bmatrix}.
\end{equation}

With the graph structure $\mathbf{S}$, the GNN refines node features over $L$ layers using a graph structure biased attention mechanism. This combines learned attention with our pre-computed similarity matrix. The feature $\mathbf{f}_i$ at layer $l+1$ is an aggregation of its neighbors:
\begin{equation}
\mathbf{f}_i^{(l+1)} = \sum_{j} \alpha_{ij}^{(l)} \cdot \mathbf{W}_v^{(l)} \mathbf{f}_j^{(l)},
\end{equation}
where $\mathbf{W}_v^{(l)}$ is a learnable value matrix. The attention coefficient $\alpha_{ij}^{(l)}$ fuses content based attention with the structural bias $\mathbf{S}_{ij}$:
\begin{equation}
\alpha_{ij}^{(l)} = \frac{\exp\left(\frac{(\mathbf{W}_q^{(l)} \mathbf{f}_i^{(l)})^T (\mathbf{W}_k^{(l)} \mathbf{f}_j^{(l)})}{\sqrt{d}} + \lambda_{\text{att}} \mathbf{S}_{ij}\right)}{\sum_{j'} \exp\left(\frac{(\mathbf{W}_q^{(l)} \mathbf{f}_i^{(l)})^T (\mathbf{W}_k^{(l)} \mathbf{f}_{j'}^{(l)})}{\sqrt{d}} + \lambda_{\text{att}} \mathbf{S}_{ij'}\right)}.
\end{equation}
Here, $\mathbf{W}_q^{(l)}$ and $\mathbf{W}_k^{(l)}$ are learnable query and key matrices. The hyperparameter $\lambda_{\text{att}}$ balances the learned attention against the fixed graph topology.

After $L$ layers, we obtain enhanced node features $\{\mathbf{f}_{\text{img},i}^{(L)}\}$ and $\{\mathbf{f}_{\text{txt},j}^{(L)}\}$. We aggregate these features using a simple pooling operation $\text{Pool}(\cdot)$ to get a global representation for each modality:
\begin{equation}
\mathbf{z}_{\text{img}} = \text{Pool}(\{\mathbf{f}_{\text{img},i}^{(L)}\}_{i=1}^{n_r}),
\end{equation}
\begin{equation}
\mathbf{z}_{\text{txt}} = \text{Pool}(\{\mathbf{f}_{\text{txt},j}^{(L)}\}_{j=1}^{n_c}),
\end{equation}
where $\mathbf{z}_{\text{img}}, \mathbf{z}_{\text{txt}} \in \mathbb{R}^{d}$. The final fused feature representation $\mathbf{g}_k$ for user $k$ is the concatenation of these pooled features:
\begin{equation}
\mathbf{g}_k = [\mathbf{z}_{\text{img}}; \mathbf{z}_{\text{txt}}] \in \mathbb{R}^{D_{\text{sem}}},
\end{equation}
where $D_{\text{sem}} = 2d$. This vector is then fed into the beamforming network.

To train this GNN and address model heterogeneity, we use a
quality-aware contrastive learning objective. We define a quality
metric $q \in [0,1]$ based on the feature's extraction depth:
\begin{equation}
q_{\text{img},i} = \text{depth}_{\text{img},i}/L_{\text{max}}, \quad q_{\text{txt},j} = \text{depth}_{\text{txt},j}/L_{\text{max}}.
\end{equation}

The intra-modal contrastive loss, $\mathcal{L}_{\text{intra}}$, encourages similar tokens within a kNN neighborhood to have aligned representations, while pushing dissimilar tokens apart. This loss is weighted by the quality $q$, giving higher quality tokens stronger influence. For visual tokens:
\begin{equation}
\mathcal{L}_{\text{intra,img}} = -\frac{1}{n_r} \sum_{i=1}^{n_r} \log \frac{\sum_{j \in \mathcal{N}_{knn}(i)} q_{\text{img},j} \cdot \exp(\text{sim}(\mathbf{f}_{\text{img},i}^{(L)}, \mathbf{f}_{\text{img},j}^{(L)}))}{\sum_{k=1}^{n_r} q_{\text{img},k} \cdot \exp(\text{sim}(\mathbf{f}_{\text{img},i}^{(L)}, \mathbf{f}_{\text{img},k}^{(L)}))},
\end{equation}
where $\text{sim}(\cdot, \cdot)$ is the cosine similarity. The loss for text tokens, $\mathcal{L}_{\text{intra,txt}}$, is computed similarly:
\begin{equation}
\mathcal{L}_{\text{intra,txt}} = -\frac{1}{n_c} \sum_{i=1}^{n_c} \log \frac{\sum_{j \in \mathcal{N}_{knn}(i)} q_{\text{txt},j} \cdot \exp(\text{sim}(\mathbf{f}_{\text{txt},i}^{(L)}, \mathbf{f}_{\text{txt},j}^{(L)}))}{\sum_{k=1}^{n_c} q_{\text{txt},k} \cdot \exp(\text{sim}(\mathbf{f}_{\text{txt},i}^{(L)}, \mathbf{f}_{\text{txt},k}^{(L)}))}.
\end{equation}
The total intra-modal loss is the sum of both:
\begin{equation}
\mathcal{L}_{\text{intra}} = \mathcal{L}_{\text{intra,img}} + \mathcal{L}_{\text{intra,txt}}.
\end{equation}

The cross-modal contrastive loss, $\mathcal{L}_{\text{cross}}$, aligns the global visual and text representations ($\mathbf{z}_{\text{img},i}, \mathbf{z}_{\text{txt},i}$) from the same sample:
\begin{equation}
\mathcal{L}_{\text{cross}} = -\frac{1}{B} \sum_{i=1}^{B} \log \frac{q_{\text{txt},i} \cdot \exp(\text{sim}(\mathbf{z}_{\text{img},i}, \mathbf{z}_{\text{txt},i}))}{\sum_{j=1}^{B} q_{\text{txt},j} \cdot \exp(\text{sim}(\mathbf{z}_{\text{img},i}, \mathbf{z}_{\text{txt},j}))},
\end{equation}
where $q_{\text{txt},i}$ is the average quality of text tokens in sample $i$. This loss pulls matched pairs together and pushes mismatched pairs apart, guided by the feature quality.

The complete contrastive objective is a weighted sum:
\begin{equation}
\mathcal{L}_{\text{contrast}} = \lambda_1 \mathcal{L}_{\text{intra}} + \lambda_2 \mathcal{L}_{\text{cross}},
\end{equation}
where $\lambda_1$ and $\lambda_2$ are hyperparameters. This objective trains the GNN to produce feature representations $\mathbf{g}_k$ that are aligned across modalities and adaptive to varying abstraction depths.

The symmetry regularization loss $\mathcal{L}_{\text{sym}}$ ensures the cross-modal similarity matrices are consistent:
\begin{equation}
\mathcal{L}_{sym} = \|\mathbf{S}_{VT} - \mathbf{S}_{TV}^T\|_F^2,
\end{equation}
where $\|\cdot\|_F$ is the Frobenius norm. This prevents asymmetric relationships from harming training.

Finally, to account for channel heterogeneity and the dynamics of wireless channels, we make the GNN framework channel-aware by modifying its initial input features. To achieve channel-aware robustness, the adapter incorporates CSI. The CSI for user $k$ consists of estimated channel vectors $\{\mathbf{\hat{h}}_k[n]\}_{n=1}^{N_c}$, where $\mathbf{\hat{h}}_k[n] \in \mathbb{C}^{N_r}$ is the estimated channel to the $N_r$ antenna server on subcarrier $n$. These vectors are flattened and encoded by an MLP into a compact embedding:
\begin{equation}
\mathbf{e}_{\text{channel},k} = \text{MLP}(\text{vec}(\mathbf{\hat{h}}_k[n]_{n=1}^{N_c})) \in \mathbb{R}^{d_{\text{emb}}},
\end{equation}
where $\text{vec}(\cdot)$ is the vectorization operator and $d_{\text{emb}}$ is the embedding dimension. In our final model, this channel embedding is concatenated with each raw feature to create the actual initial node features, which we denote as $\mathbf{f'}^{(0)}$:
\begin{equation}
\begin{aligned}
\mathbf{f'}_{\text{img},i}^{(0)} &= [\mathbf{r}_i; \mathbf{e}_{\text{channel},k}] \in \mathbb{R}^{d+d_{\text{emb}}}, \quad i = 1, \ldots, n_r, \\
\mathbf{f'}_{\text{txt},j}^{(0)} &= [\mathbf{c}_j; \mathbf{e}_{\text{channel},k}] \in \mathbb{R}^{d+d_{\text{emb}}}, \quad j = 1, \ldots, n_c,
\end{aligned}
\end{equation}
where $[\cdot; \cdot]$ denotes concatenation. This initialization makes every node in the graph aware of the user's channel conditions. This channel-aware feature $\mathbf{f'}^{(0)}$ is then processed by the same GNN modeling logic, resulting in a new channel-aware similarity graph $\mathbf{S'}$ and a final robust feature representation $\mathbf{g'}_k\in \mathbb{R}^{D'_{\text{sem}}}, D'_{\text{sem}} = 2(d+d_{emb})$. This $\mathbf{g'}_k$ is robust to channel distortions, aligned across modalities, and adaptive to varying abstraction depths, and is the vector ultimately fed into the following beamforming network.

\subsection{Beamforming Network}
Given the feature $\mathbf{g'}_k$ from the adapter for each user $k$, the beamforming network maps it to a frequency-domain symbol vector $\mathbf{s}_k = [s_k[1], \ldots, s_k[N_c]]^T \in \mathbb{C}^{N_c}$ for transmission. We adopt a hybrid approach that combines data-driven and model-driven strategies to balance adaptability and robustness.

\subsubsection{Model-Driven OTA Alignment Symbol Mapping}

To realize constructive OTA aggregation under heterogeneous fading and imperfect CSI, we adopt an MMSE-inspired model-driven symbol mapping
to mitigate channel distortion in OTA aggregation\cite{hu2020iterative,shi2023robust}. The core idea is to represent the multi-antenna reception by a learnable linear combiner, form an equivalent scalar channel per client and subcarrier, and obtain a closed-form robust pre-equalizer.

The equivalent scalar channel $h_{\text{eff},k}[n]$ is formed by applying a reference combining vector $\mathbf{v}[n]\in\mathbb{C}^{N_r\times 1}$ to the physical channel vector $\mathbf{h}_k[n]\in\mathbb{C}^{N_r}$. To resolve the inherent scale ambiguity in transmitter optimization, this combining vector is constrained to have a unit norm, i.e., $\left\|\mathbf{v}[n]\right\|=1$. The effective channel is thus defined as:
\begin{equation}
h_{\text{eff},k}[n]=\mathbf{v}[n]^H\mathbf{h}_k[n]\in\mathbb{C}.
\end{equation}
The combining vector $\mathbf{v}[n]$ is a robust pre-equalizer parameter and is jointly optimized with all other network parameters during training.

For feature projection, the GNN provides a channel-aware feature vector $\mathbf{g'}_k\in\mathbb{R}^{D'_{\text{sem}}}$ for user $k$. We project it to a complex symbol per subcarrier via a learnable vector $\mathbf{w}_n\in\mathbb{C}^{D'_{\text{sem}}\times 1}$:
\begin{equation}
\tilde{s}_k[n]=\mathbf{w}_n^H\mathbf{g'}_k\in\mathbb{C}.
\end{equation}
The set $\{\mathbf{w}_n\}_{n=1}^{N_c}$ is trained end-to-end to adapt to the frequency-selective features.

After receiving combining, the aggregated scalar signal at subcarrier $n$ reads
\begin{equation}
\hat{y}[n]=\frac{1}{\sqrt{K}}\sum_{k=1}^K h_{\text{eff},k}[n]\,s_k[n]+ \tilde{z}[n],
\end{equation}
where $\tilde{z}[n]\sim\mathcal{CN}(0,\tilde{\sigma}_n^2)$ with $\tilde{\sigma}_n^2=  \sigma_n^2\|\mathbf{v}[n]\|^2$ is the effective post-combining noise variance. The model-driven branch aims to compensate for
channel-induced phase rotation and support coherent
feature aggregation.




To promote coherent aggregation under imperfect CSI,
we use a regularized channel-conjugate pre-equalizer.
Let $\hat h_{\mathrm{eff},k}[n]
=\mathbf{v}[n]^{H}\hat{\mathbf{h}}_k[n]$
denote the estimated effective channel, and let
$\epsilon_k[n]
=\mathbb{E}[|h_{\mathrm{eff},k}[n]
-\hat h_{\mathrm{eff},k}[n]|^2]$
denote its estimation-error variance.
The coefficient for user $k$ on subcarrier $n$ is defined as:
\begin{equation}
\label{eq:robust_pre_eq}
\begin{aligned}
F_k[n] &= \frac{\sqrt{K} \hat{h}_{\text{eff},k}^*[n]}{\sum_{j=1}^K |\hat{h}_{\text{eff},j}[n]|^2 + \sum_{j=1}^K \epsilon_j[n] + \lambda\,\tilde{\sigma}_n^2} \\
&\approx \frac{\sqrt{K} \hat{h}_{\text{eff},k}^*[n]}{\sum_{j=1}^K |\hat{h}_{\text{eff},j}[n]|^2 + K \sigma_e^2 + \lambda\,\tilde{\sigma}_n^2},
\end{aligned}
\end{equation}
where the numerator, $\hat{h}_{\text{eff},k}^*[n]$, performs phase conjugation to coherently align the transmitted symbol with the effective channel for constructive combining. The denominator is a robust normalization factor that balances the total multi-user channel power, $\sum_{j=1}^K |\hat{h}_{\text{eff},j}[n]|^2$, against the regularized noise power, $\lambda\,\tilde{\sigma}_n^2$, and the aggregated channel uncertainty. $\lambda>0$ controls the strength of
noise-dependent regularization. We assume the channel estimation error follows an independent and identically distributed (i.i.d.) complex Gaussian distribution with variance $\sigma_e^2$
, where simplifies the per-user error variance to a uniform scalar $\epsilon_j[n] \approx \sigma_e^2, \forall j$, yielding the tractable aggregated uncertainty $K \sigma_e^2$. This MMSE-based structure prevents excessive power allocation to users with weak channels and reduces to the standard form when CSI is perfect (i.e., $\sigma_e^2 = 0$).

The model-driven symbol for user $k$ on subcarrier $n$ is then defined as:
\begin{equation}
s_k^{(m)}[n] = F_k[n] \cdot \tilde{s}_k[n].
\end{equation}

\subsubsection{Data-Driven Feature-Aware Beamforming}

To complement the model-driven approach with enhanced adaptability, we design a data-driven beamforming network that learns the end-to-end mapping from feature vectors to transmission symbols via a Transformer architecture \cite{vaswani2017attention,han2022survey}.

For discriminative weighting, we apply a learned attention mechanism to emphasize important dimensions of the feature:
\begin{equation}
\begin{aligned}
    \boldsymbol{\alpha}_k &= \sigma(\mathbf{W}_{\alpha} \mathbf{g'}_k + \mathbf{b}_{\alpha}) \in [0,1]^{D'_{\text{sem}}}, \\
    \tilde{\mathbf{g}}_k &= \boldsymbol{\alpha}_k \odot \mathbf{g'}_k,
\end{aligned}
\end{equation}
where $\mathbf{W}_{\alpha} \in \mathbb{R}^{D'_{\text{sem}} \times D'_{\text{sem}}}$ 
and $\mathbf{b}_{\alpha} \in \mathbb{R}^{D'_{\text{sem}}}$ are learnable parameters, $\sigma(\cdot)$ is the sigmoid activation function, and $\odot$ denotes element-wise multiplication.

Next, linear embedding and positional encoding are applied. The weighted feature vector is linearly embedded to the model dimension:
\begin{equation}
\mathbf{e}_k = \mathbf{W}_{\text{emb}} \tilde{\mathbf{g}}_k + \mathbf{b}_{\text{emb}}, \quad \mathbf{e}_k \in \mathbb{R}^{d_{\text{model}}},
\end{equation}
where $\mathbf{W}_{\text{emb}} \in \mathbb{R}^{d_{\text{model}} \times D'_{\text{sem}}}$ 
and $\mathbf{b}_{\text{emb}} \in \mathbb{R}^{d_{\text{model}}}$ are learnable parameters. To incorporate frequency information, we expand embedding to subcarriers with sinusoidal positional encoding:
\begin{equation}
\mathbf{E}_k[n] = \mathbf{e}_k + \mathbf{PE}[n], \quad n = 1, \ldots, N_c,
\end{equation}
where $\mathbf{PE}[n] \in \mathbb{R}^{d_{\text{model}}}$ is the fixed positional encoding vector from the original Transformer architecture.

The sequence $\{\mathbf{E}_k[n]\}_{n=1}^{N_c}$ is then processed by a Transformer encoder to capture inter-subcarrier dependencies:
\begin{equation}
\mathbf{Z}_k = \text{TransformerEncoder}(\mathbf{E}_k) \in \mathbb{R}^{N_c \times d_{\text{model}}},
\end{equation}
where each layer consists of multi-head self-attention (MSA) and a feed-forward network (FFN):
\begin{equation}
\begin{aligned}
    \mathbf{Z}_k^{(\ell)} &= \text{LayerNorm}\left(\mathbf{Z}_k^{(\ell-1)} + \text{MSA}(\mathbf{Z}_k^{(\ell-1)})\right), \\
    \mathbf{Z}_k^{(\ell)} &= \text{LayerNorm}\left(\mathbf{Z}_k^{(\ell)} + \text{FFN}(\mathbf{Z}_k^{(\ell)})\right).
\end{aligned}
\end{equation}

For complex symbol mapping, we map the contextual representation to complex-valued symbols. This is achieved by mapping the real-valued output to a 2D vector which is then interpreted as the real and imaginary parts of a complex symbol:
\begin{equation}
\begin{aligned}
    \mathbf{u}_k[n] &= \mathbf{W}_{\text{out}} \mathbf{Z}_k[n] + \mathbf{b}_{\text{out}}, \quad \mathbf{u}_k[n] \in \mathbb{R}^2, \\
    s_k^{(d)}[n] &= u_k^{(1)}[n] + j u_k^{(2)}[n],
\end{aligned}
\end{equation}
where $\mathbf{W}_{\text{out}} \in \mathbb{R}^{2 \times d_{\text{model}}}$ and $\mathbf{b}_{\text{out}} \in \mathbb{R}^2$ are learnable parameters. The complete data-driven mapping is denoted as:
\begin{equation}
\mathbf{s}_k^{(d)} = \mathcal{T}_{\text{data}}(\mathbf{g'}_k; \Theta_d) \in \mathbb{C}^{N_c},
\end{equation}
where $\Theta_d$ encompasses all learnable parameters in the data-driven branch.

\subsubsection{Hybrid Symbol Mapping}

To leverage the strengths of both data-driven adaptability and model-driven robustness, we fuse the two symbol mapping strategies via a learnable weighted sum:
\begin{equation}
\mathbf{s}_k = \beta_d \mathbf{s}_k^{(d)} + \beta_m \mathbf{s}_k^{(m)},
\end{equation}
where the weights are parameterized by learnable scalars $\theta_d, \theta_m \in \mathbb{R}$ via a softmax function:
\begin{equation}
\beta_d = \frac{\exp(\theta_d)}{\exp(\theta_d) + \exp(\theta_m)}, \quad \beta_m = \frac{\exp(\theta_m)}{\exp(\theta_d) + \exp(\theta_m)}.
\end{equation}
This softmax parameterization ensures $\beta_d + \beta_m = 1$ and $\beta_d, \beta_m \in [0,1]$. During end-to-end training, $\theta_d$ and $\theta_m$ are optimized jointly with all other network parameters, allowing the system to automatically learn the optimal balance between the two strategies based on the task objective and channel conditions.

\subsubsection{Power Normalization}

To satisfy the per-user power constraint, we normalize the fused symbol vector:
\begin{equation}
\mathbf{s}_k \leftarrow \sqrt{\frac{P_k}{\sum_{n=1}^{N_c} |s_k[n]|^2}} \cdot \mathbf{s}_k,
\end{equation}
where $P_k$ is the power budget of user $k$. This normalization is applied after hybrid fusion to ensure that the transmitted signal adheres to hardware limitations while preserving the relative amplitude and phase relationships learned by the network.

\subsection{OTA Aggregation and CI}

The symbol vectors $\mathbf{s}_k \in \mathbb{C}^{N_c}$ from all $K$ users, generated by the hybrid beamforming network, are transmitted simultaneously. These signals propagate through their respective multipath channels $\{\mathbf{h}_k[n]\}$ and are naturally superimposed at the server's $N_r$-antenna array. As derived in \eqref{OTA_agg} and \eqref{OTA_receive}, the server receives the complete space-frequency signal matrix $\mathbf{Y} \in \mathbb{C}^{N_c \times N_r}$, where the $n$-th row $\mathbf{y}[n]^T$ contains the $N_r$-dimensional vector from all users on that subcarrier.

This entire received signal matrix $\mathbf{Y}$ is treated as the input wavefront to SIM. In practice, we preprocess $\mathbf{Y}$ to match the SIM input layer size via interpolation. We then vectorize the aperture samples to form a single input vector $\mathbf{y}_{\text{vec}} \in \mathbb{C}^{M}$, where $M = N_c \cdot N_r$:
\begin{equation}
    \mathbf{y}_{\text{vec}} = \text{vec}(\mathbf{Y}).
\end{equation}
The receiver SIM is modeled as one comprehensive, trainable linear transformation matrix $\mathbf{\Theta}_{\text{SIM}} \in \mathbb{C}^{C \times M}$, where $C$ is the number of classes. This single matrix is physically realized by the SIM’s multi-layer structure:
\begin{equation}
    \mathbf{\Theta}_{\text{SIM}} = \mathbf{Z}^{(L_{rx})}\mathbf{W}^{(L_{rx})} \cdots \mathbf{Z}^{(1)}\mathbf{W}^{(1)},
\end{equation}
where $\mathbf{W}^{(l)}$ are fixed physical propagation matrices, and $\mathbf{Z}^{(l)}$ are diagonal matrices representing the trainable meta-atoms of the $l$-th layer. Concretely, each $\mathbf{W}^{(l)}$ implements free-space diffraction between two parallel planes via the Rayleigh--Sommerfeld diffraction equation, while $\mathbf{Z}^{(l)}=\mathrm{diag}(A^{(l)}\odot e^{j\phi^{(l)}})$ with $0\le A^{(l)}_n<1$ and $0\le \phi^{(l)}_n<2\pi$. The stack is fully differentiable and can be trained by back-propagated gradients to $\{A^{(l)},\phi^{(l)}\}$.

This unified SIM model provides the flexibility to process signals from both the data-driven and model-driven transmitter paths. The complete received 
space-frequency signal matrix $\mathbf{Y}$ contains the naturally aggregated contributions from all users, each pre-processed by their respective transmit strategies. The SIM, through its end-to-end trained transformation $\mathbf{\Theta}_{\text{SIM}}$, learns to perform the necessary spatial-frequency combining and feature classification in a single physical wave propagation. 
This joint optimization of the transmit side networks
and the receive side SIM enables the entire system to discover an optimal end-to-end mapping from heterogeneous user features to classification outcomes, all executed at the speed of light with minimal energy consumption.

The SIM processes the input vector $\mathbf{y}_{\text{vec}}$ to produce a final output signal $\mathbf{p}_{\text{out}} \in \mathbb{C}^{C}$ at $C$ designated output points:
\begin{equation}
    \mathbf{p}_{\text{out}} = \mathbf{\Theta}_{\text{SIM}} \cdot \mathbf{y}_{\text{vec}}.
\end{equation}
The energy at each of the $C$ output ports is then detected, and this energy value $p_c$ serves as the logit for the corresponding class $c$:
\begin{equation}
    p_c = |(p_{\text{out}})_c|^2, \quad \forall c \in \{1, \ldots, C\}.
\end{equation}
For the multi-label task, these logits are passed through a sigmoid function to obtain an independent probability $\hat{p}_c$ for each class:
\begin{equation}
    \hat{p}_c = \sigma(p_c) = \frac{1}{1 + \exp(-p_c)}, \quad \forall c \in \{1, \ldots, C\}.
\end{equation}

The final task loss is the Binary Cross-Entropy (BCE) loss between the predicted probabilities $\hat{\mathbf{p}}$ and the ground truth label vector $\mathbf{y}_{\text{true}}$:
\begin{equation}
    \mathcal{L}_{\text{task}} = -\sum_{c=1}^{C} \left( (y_{\text{true}})_c \log(\hat{p}_c) + (1-(y_{\text{true}})_c) \log(1-\hat{p}_c) \right).
\end{equation}
The total loss function for the entire system is a weighted sum of the task loss and the self-supervised losses from the GNN:
\begin{equation}
    \mathcal{L}_{\text{total}} = \mathcal{L}_{\text{task}} + \lambda_{\text{contrast}} \mathcal{L}_{\text{contrast}} + \lambda_{\text{sym}} \mathcal{L}_{\text{sym}}.
\end{equation}

\begin{algorithm}[t]
\caption{Training Process of SIM-CI}
\label{alg:framework}
\begin{algorithmic}[1]
\REQUIRE Multi-modal data $\{\mathcal{D}_k\}_{k=1}^K$, channel info $\{\mathbf{h}_k, \hat{\mathbf{h}}_k\}_{k=1}^K$, labels $\mathbf{y}_{\text{true}}$

\STATE Initialize Parameters GNN $\Theta_{\text{GNN}}$, Beamformer $\Theta_{\text{beam}}$, SIM $\Theta_{\text{SIM}}$

\FOR{each training iteration}
    \FOR{each user $k = 1$ to $K$}
        \STATE Extract tokens $\{\mathbf{r}_i^k\}, \{\mathbf{c}_j^k\}$ and $\text{depth}^k$ from $\mathcal{D}_k$

        \STATE $\mathbf{g}_k, \mathcal{L}_{\text{contrast}}^k, \mathcal{L}_{\text{sym}}^k \gets \text{GNN}(\{\mathbf{r}_i^k\}, \{\mathbf{c}_j^k\}, \hat{\mathbf{h}}_k, \text{depth}^k)$

        \STATE $\mathcal{L}_{\text{client}} \gets \mathcal{L}_{\text{client}} + \mathcal{L}_{\text{contrast}}^k + \mathcal{L}_{\text{sym}}^k$
        
        \STATE $\mathbf{s}_k \gets \text{Beamformer}(\mathbf{g}_k, \hat{\mathbf{h}}_k)$
        \STATE $\mathbf{s}_k \gets \sqrt{P_k / \|\mathbf{s}_k\|^2} \cdot \mathbf{s}_k$
    \ENDFOR
    
    \STATE $\mathbf{Y} \gets \text{OTA}(\{\mathbf{s}_k\}_{k=1}^K, \{\mathbf{h}_k\}_{k=1}^K, \sigma_n^2)$

    \STATE $\mathbf{y}_{\text{vec}} \gets \text{vec}(\mathbf{Y})$
    \STATE $\mathbf{p}_{\text{out}} \gets \Theta_{\text{SIM}} \cdot \mathbf{y}_{\text{vec}}$
    \STATE $\{p_c\} \gets \{|(p_{\text{out}})_c|^2\}_{c=1}^C$
    \STATE $\hat{\mathbf{p}} \gets \sigma(\{p_c\})$
    
    \STATE $\mathcal{L}_{\text{task}} \gets \text{BCE}(\hat{\mathbf{p}}, \mathbf{y}_{\text{true}})$
    \STATE $\mathcal{L}_{\text{total}} \gets \mathcal{L}_{\text{task}} + \lambda_{\text{client}} \mathcal{L}_{\text{client}}$
    \STATE Update $\Theta_{\text{GNN}}, \Theta_{\text{beam}}, \Theta_{\text{SIM}}$ using $\nabla \mathcal{L}_{\text{total}}$
\ENDFOR

\RETURN $\Theta_{\text{GNN}}, \Theta_{\text{beam}}, \Theta_{\text{SIM}}$
\end{algorithmic}
\end{algorithm}

\section{Performance Results and Analysis} \label{Sec:evaluation}

In this section, we present the simulation settings and illustrate the superior performance of the proposed SIM-CI.

\subsection{Experimental Setting}

\textbf{Dataset:} We evaluate our SIM-CI framework on the COCO2017 dataset \cite{lin2014microsoft}, a large-scale object detection and classification benchmark containing 118,287 training images and 5,000 validation images across 80 object categories. The dataset is used for multi-label classification tasks, and its multi-modal characteristics make it well-suited for evaluating multi-user collaborative inference scenarios. To further evaluate modality heterogeneity, we additionally use SpokenCOCO \cite{hsu2021text}, which provides spoken captions associated with COCO images. For the tri-modal experiment, the speech samples are matched with their corresponding image and text samples, providing aligned image, text, and speech observations for the same semantic content.

\textbf{System and Channel Model:} We simulated a multi-user OFDM uplink system with a carrier frequency $f_c = 2.6$ GHz, subcarrier spacing $\Delta f = 60$ kHz, and $N_c = 64$ subcarriers. The system comprises $K=4$ single-antenna users and a receiver SIM equipped with an $N_r = 64$ ($8 \times 8$ UPA) antenna array. Each user's channel is modeled as a multi-path Rayleigh fading channel with $L_k=2$ propagation paths, with angles of arrival randomly distributed. The noise power spectral density was set to $N_0 = -174$ dBm/Hz.

\textbf{Network Architecture:} Feature extraction employs pre-trained deep learning encoders. For the visual modality, we use Faster R-CNN with a ResNet-101 backbone to extract region features of dimension 2048, extracting $n_r = 36$ regions per image. For the textual modality, we employ the BERT-base-uncased model to generate token-level features of dimension 768. Features from both modalities are mapped to a unified dimension $d = 1024$ through linear projection layers.

Our designed cross-modal GNN adapter adopts $L = 1$ architecture. For similarity matrix construction, intra-modal similarities are computed using a k-nearest neighbor constrained RBF kernel with $k = 10$ and bandwidth $\sigma = 1.0$. Cross-modal similarity computation employs top-K selection with $k_{\text{cross}} = 10$. The attention scaling parameter in the message passing mechanism is set to $\lambda_{\text{att}} = 1.5$, and global feature aggregation uses max pooling. The data-driven branch of the hybrid beamforming network uses a Transformer encoder with $d_{model} = 256$, 4 encoder layers, and 8 attention heads.

\textbf{Training Configuration:} The experiments were executed in the Lancaster University HEC GPU cluster using NVIDIA V100 (32 GB) and L40 (48 GB) GPU nodes through the gpu-short / gpu-medium queues. The batch size is set to $B = 128$. For the adapter loss, the intra-modal and cross-modal components are weighted by $\lambda_1 = 0.5$ and $\lambda_2 = 0.5$, respectively. The total loss function balances the task loss with contrastive and symmetry losses using $\lambda_{\text{contrast}} = 0.5$ and $\lambda_{\text{sym}} = 0.5$. We use the Adam optimizer with a learning rate of $1 \times 10^{-4}$.

\textbf{Evaluation Metrics:} We use mean Average Precision (mAP) as the primary evaluation metric to measure the accuracy of multi-label classification.

\subsection{SIM-CI Accuracy and Computational Overhead}

In this section, we evaluate the inference accuracy and computational overhead of our proposed SIM-CI framework. Our approach integrates a multi-modal adapter with a SIM-based physical classifier. We compare its performance against five key baselines chosen to represent different computational paradigms and ablation scenarios. 

As the ideal digital-domain counterpart, we use BLIP-2 \cite{li2023blip}, which performs multi-modal fusion and classification entirely in the digital domain. To demonstrate the value of multi-modal fusion, we compare against single-modality-only baselines. For the SIM-based single-modality ablations, we employ SIM-TSC \cite{huang2024stacked} for image-only inference. For the text-only imference, we create a comparable baseline based on Adapter-BERT \cite{houlsby2019parameter}, where we replace its classifier with our simulated SIM network to process the adapter processed text tokens. For standard digital-domain classifiers, we benchmark against ViT \cite{dosovitskiy2020image} for image classification and RoBERTa \cite{liu2019roberta} for text classification.

To quantify the performance and efficiency of the receiver, we designed a digital simulation experiment. We constructed an $L$-layer SIM model and compared it against an $L$-layer standard DNN (MLP) baseline of equivalent depth. The fundamental difference between them lies in the parameter space. A DNN is an ideal model possessing $O(L \cdot N^2)$ fully trainable, dense weight matrices ($W^l$) and bias vectors ($b^l$). In contrast, the SIM model is strictly constrained by physical laws, with its trainable parameters limited to only the $O(L \cdot N)$ diagonal response matrices $Z^l$.

\begin{figure}[t]
    \centering
        \includegraphics[width=0.45\textwidth]{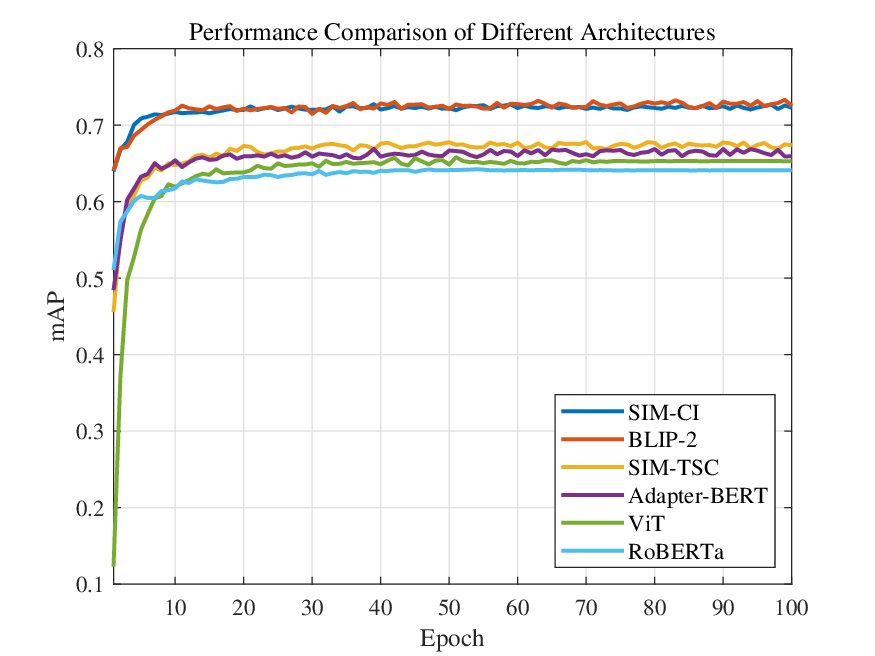}
    \caption{Inference accuracy (mAP) vs. training epoch. Performance comparison of the proposed SIM-CI framework against multi-modal (BLIP-2) and single-modality (SIM-TSC, Adapter-BERT, ViT, RoBERTa) baselines.}
    \label{fig:SIM}
\end{figure}

Figure \ref{fig:SIM} shows the mAP performance during training for all six schemes . The multi-modal schemes (SIM-CI and BLIP-2) clearly outperform the single-modality ones (SIM-TSC, Adapter-BERT, ViT, RoBERTa), validating the need for fusing heterogeneous data. Moreover, our proposed SIM-CI almost perfectly matches the ideal digital BLIP-2 baseline. Both converge at the same rate and reach nearly identical final mAP scores, 0.733 for BLIP-2 and 0.723 for SIM-CI. The figure thus shows that the SIM's physical structure has enough model capacity to match the ideal digital accuracy.

Table~\ref{tab:performance_complete} compares six methods across computational energy, spectral efficiency, and computational latency.

\textbf{Computational Energy.} The total energy covers both transmitter processing (feature extraction, adapter, beamforming) and receiver operations (SIM's physical detection vs. digital inference). In the multi-modal case (comparing SIM-CI and BLIP-2), both methods use the same adapter, so the energy gap (0.0060K vs 0.0080K) is due only to the receiver. The BLIP-2 digital baseline needs heavy digital processing, while SIM-CI uses simple energy detection. For single-modality (\emph{e.g.}, comparing SIM-TSC to ViT), the SIM-based method has an adapter while the DNN-based one does not, but SIM-TSC still has lower total energy (\emph{e.g.}, 0.0042K vs 0.0045K for image). This is because the DNN's receiver cost is much larger than the SIM's adapter cost. As $M$ increases, the digital layers energy grows linearly because its receiver network scales with the input size. SIM's cost stays constant, showing its $\mathcal{O}(1)$ receiver scaling.

\textbf{Spectral Efficiency.} Digital-domain methods (BLIP-2, ViT, RoBERTa) use TDMA, which needs $K$ separate time slots. This means their spectral efficiency drops as the user count $K$ increases, \emph{i.e.}, proportional to $1/K$. In contrast, SIM-based methods (SIM-CI, SIM-TSC, Adapter-BERT) use OTA aggregation, allowing all users to transmit at the same time. As a result, its efficiency stays constant regardless of $K$. When scaling $M$, both methods improve from array gain by logarithmic growth, but SIM's better physical-layer beamforming leads to a steeper improvement.

\textbf{Computational Latency.} We model latency by assuming that a single floating-point operation (FLOP) takes 1 nanosecond on GPUs, a neural network layer's computation time scales linearly with its FLOPs, and the transmission duration for one OFDM symbol is 1 millisecond per time slot. Physical wave propagation delay is negligible compared to these processing delays. Because digital-domain methods (BLIP-2, ViT, RoBERTa) process users serially, their latency grows linearly with $K$. In contrast, SIM's parallel OTA keeps its latency constant. In the single-modality case  (\emph{e.g.}, SIM-TSC vs. ViT), the SIM's adapter adds almost no latency compared to the digital receiver cost. That's why the SIM-based approach is still faster, even with the extra adapter. Finally, as $M$ scales, the digital layer latency grows linearly, while SIM's stays constant.

The SIM physical layer performs inference at the speed of light with $\mathcal{O}(1)$ receiver complexity, offering a fundamentally low-energy and low-latency solution. Based on Table~\ref{tab:performance_complete}, SIM-CI reduces computational energy by 25\%, reduces spectral efficiency by 75\%, and reduces computational latency by 76.1\% compared to the BLIP-2 digital baseline under our simulation design, where $K=4$ and $M=4096$.

\begin{table*}[t]
\centering
\caption{Overhead Analysis: Computational Energy, Spectral Efficiency, and Computational Latency. The table presents overhead functions versus the number of users ($K$) and the receiver input dimension ($M$).}
\label{tab:performance_complete}
\renewcommand{\arraystretch}{1.35}
\resizebox{\textwidth}{!}{%
\begin{tabular}{lcccccc}
\hline\hline
\textbf{Method} &
\multicolumn{2}{c}{\textbf{Computational Energy (J)}} &
\multicolumn{2}{c}{\textbf{Spectral Efficiency  (bit/s/Hz)}} &
\multicolumn{2}{c}{\textbf{Computation Latency (ms)}} \\
\cline{2-3} \cline{4-5} \cline{6-7}
 & vs $K$ & vs $M$
 & vs $K$ & vs $M$
 & vs $K$ & vs $M$ \\
\hline

SIM-CI &
$0.0060K$ &
$0.024$ &
$4.00$ &
$4\left(1+0.80\ln\frac{M}{4096}\right)$ &
$13.10$ &
$13.10$ \\

BLIP-2 \cite{li2023blip} &
$0.0080K$ &
$0.0299+5\times10^{-7}M$ &
$\frac{4}{K}$ &
$\log_2\left(1+\frac{M}{4096}\right)$ &
$13.70K$ &
$52.71+0.00051M$ \\

SIM-TSC \cite{huang2024stacked} &
$0.0042K$ &
$0.0168$ &
$2.57$ &
$2.57\left(1+0.80\ln\frac{M}{4096}\right)$ &
$10.50$ &
$10.50$ \\

Adapter-BERT \cite{houlsby2019parameter} &
$0.0024K$ &
$0.0096$ &
$1.43$ &
$1.43\left(1+0.80\ln\frac{M}{4096}\right)$ &
$8.70$ &
$8.70$ \\

ViT \cite{dosovitskiy2020image} &
$0.0045K$ &
$0.0167+3\times10^{-7}M$ &
$\frac{2.57}{K}$ &
$\frac{2.57}{4}\log_2\left(1+\frac{M}{4096}\right)$ &
$11.00K$ &
$42.32+0.00041M$ \\

RoBERTa \cite{liu2019roberta} &
$0.0025K$ &
$0.009+2\times10^{-7}M$ &
$\frac{1.43}{K}$ &
$\frac{1.43}{4}\log_2\left(1+\frac{M}{4096}\right)$ &
$9.20K$ &
$35.41+0.00034M$ \\

\hline\hline
\end{tabular}%
}
\end{table*}

\subsection{Robustness to Channel Heterogeneity}

To validate the robustness and effectiveness of the proposed framework against channel heterogeneity, we conducted an ablation study. This study evaluates the performance of the channel-aware GNN and the hybrid beamforming network.


\begin{figure}[t]
    \centering
        \centering
        \includegraphics[width=0.5\textwidth]{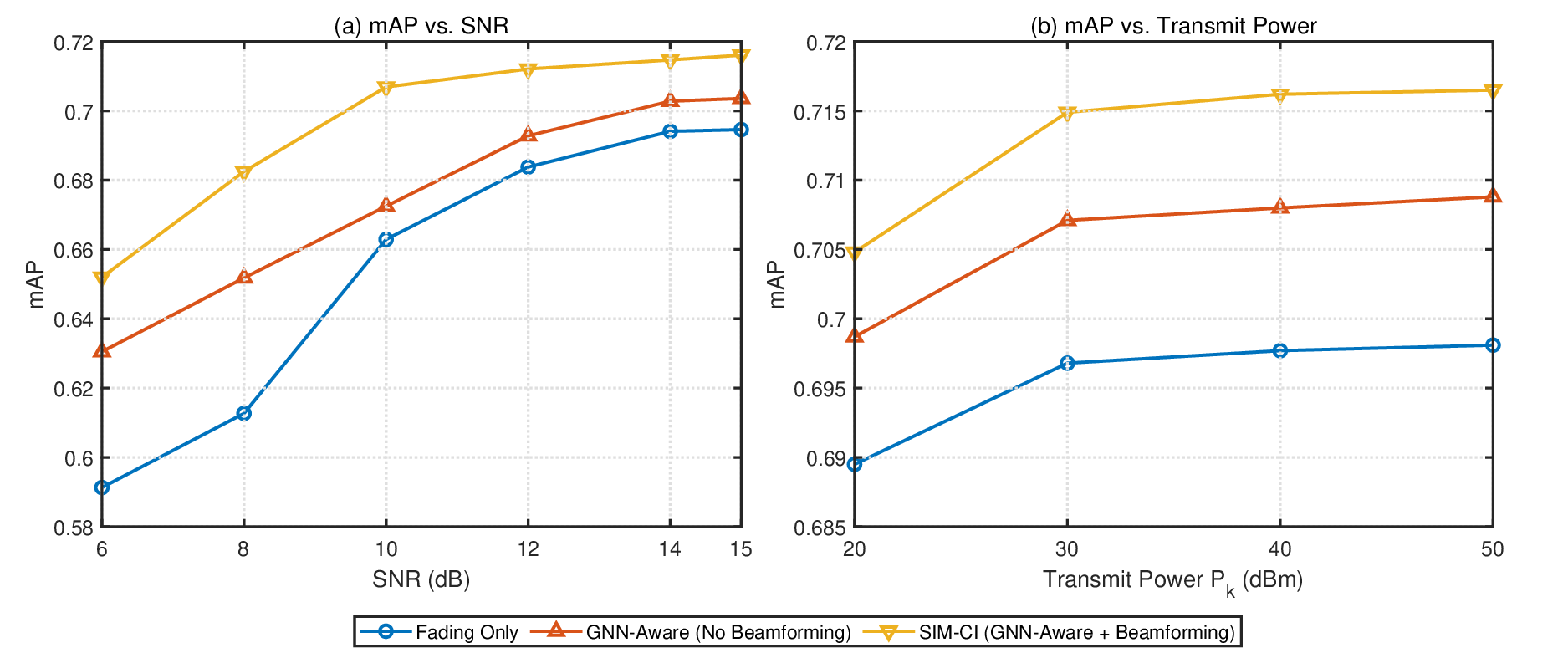}
    \caption{Ablation study on the channel robustness components of the SIM-CI framework. (a) mAP versus Signal-to-Noise Ratio (SNR). (b) mAP versus transmit power ($P_k$).}
    \label{fig:CSI}
    \centering
\end{figure}





We compared the following three configurations:
\begin{itemize}
    \item Fading Only: In this scheme, the GNN does not receive CSI as input, \emph{i.e.}, $\mathbf{e}_{\text{channel},k}$ is removed. The beamforming network degenerates to a data-driven-only Transformer ($\mathbf{s}_k = \mathbf{s}_k^{(d)}$) that is also channel-unaware.
    \item GNN-Aware (No Beamforming): Here, the GNN receives CSI as input, allowing it to learn implicit robustness in the feature space. However, it still uses the same channel-unaware, data-driven-only beamformer as the baseline.
    \item SIM-CI (GNN-Aware + Beamforming): This is the complete framework proposed in this paper. The GNN is channel-aware, and the hybrid beamforming network performs robust pre-equalization.
\end{itemize}

Figure \ref{fig:CSI}(a) shows the mAP performance as a function of the Signal-to-Noise Ratio (SNR), from 6 dB to 15 dB. As expected, the performance of all three schemes improves with increasing SNR.  A consistent performance hierarchy is observed across the entire SNR range, where SIM-CI shows the best performance, followed by the GNN-Aware scheme and the Baseline. The performance gap between the GNN-Aware scheme and fading-only (e.g., 0.672 vs. 0.662 mAP at 10 dB SNR) demonstrates the benefit of making the GNN channel-aware. This allows the GNN to project features into a subspace that is implicitly more robust to channel distortions. The additional performance gain of SIM-CI over the GNN-Aware scheme (e.g., 0.706 mAP at 10 dB) underscores the necessity of explicit signal-level pre-compensation. 

Figure \ref{fig:CSI}(b) plots mAP against user transmit power $P_k$, shows the system's performance bottleneck. The fading-only scheme's performance saturates quickly at $P_k = 30$ dBm, plateauing at an mAP of approximately 0.698. This indicates the system becomes limited by channel distortion, not noise. Once in this state, performance no longer improves with higher power, as the signal misalignment caused by fading remains uncorrected. The GNN-Aware scheme lifts this performance ceiling to $\approx 0.708$, showing that its implicit feature compensation can mitigate some distortion. Finally, SIM-CI achieves the highest plateau ($\approx 0.717$). This demonstrates that the hybrid beamforming network is essential. By actively performing amplitude pre-equalization and phase pre-alignment, it mitigates the channel distortion, breaking the bottleneck and enabling accurate physical-layer feature aggregation.

\subsection{Robustness to Model-Depth Heterogeneity}

\begin{figure}[t]
    \centering
        \centering
        \includegraphics[width=0.45\textwidth]{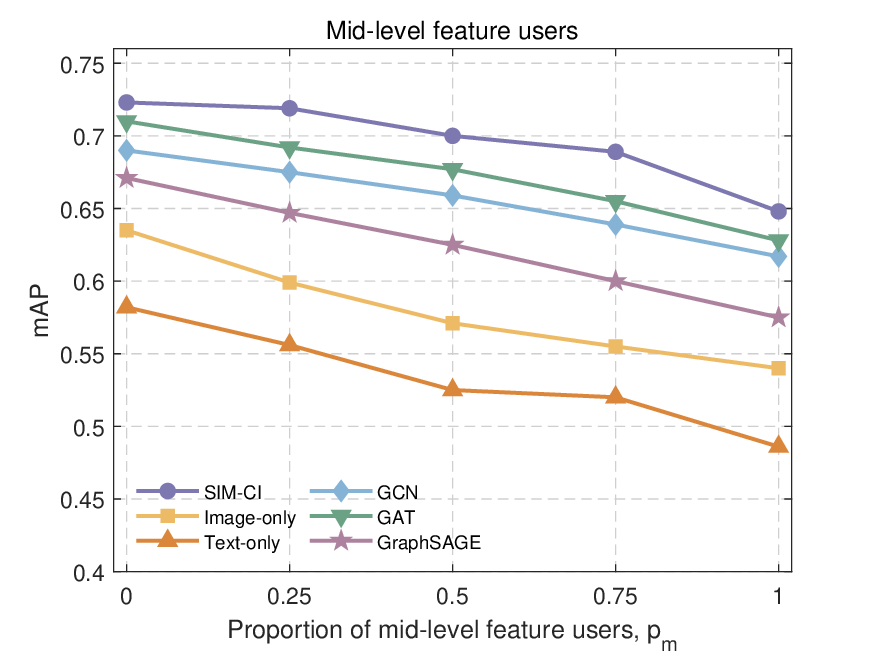}
    \caption{Robustness to model-depth heterogeneity with increasing proportion of mid-level feature users.}
    \label{fig:mid}
    \centering
\end{figure}

\begin{figure}[t]
    \centering
        \centering
        \includegraphics[width=0.45\textwidth]{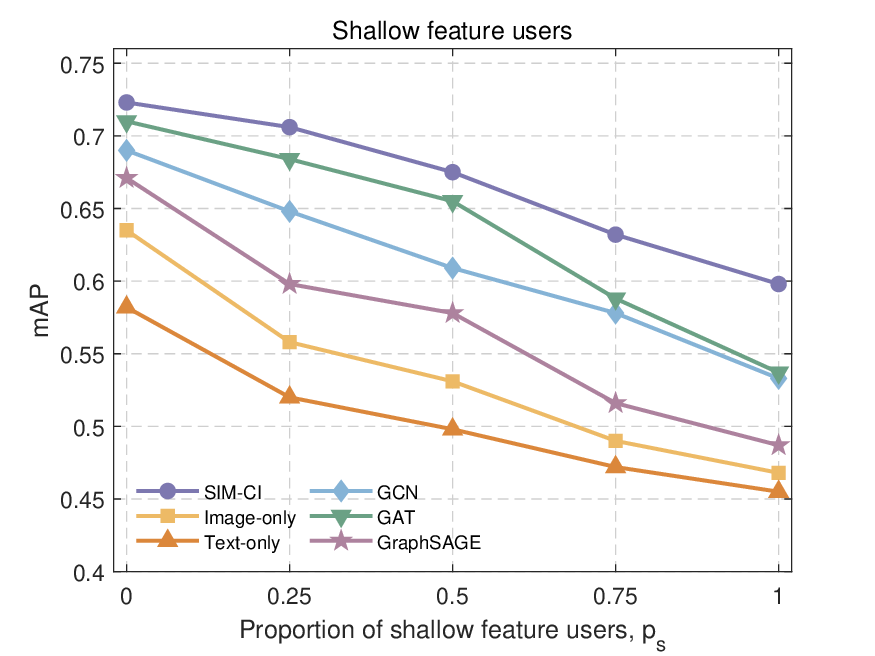}
    \caption{Robustness to model-depth heterogeneity with increasing proportion of shallow feature users.}
    \label{fig:shallow}
    \centering
\end{figure}

Fig. \ref{fig:mid} and Fig. \ref{fig:shallow} show the influence of model-depth heterogeneity on SIM-CI. Users with
different computational power extract features at different depths, resulting in varied
feature quality. We construct shallow, mid, and deep features directly from different
network depths, image features from the res3, res4, and res5 stages of a Faster R-CNN with
a ResNet-101 backbone, and text features from layers 3, 6, and 12 of BERT-base. Training
uses all three depths jointly, so the adapter learns to align these different abstraction
levels.

At test time we fix $K=4$ users and vary the proportion of users providing mid-level
features, $p_m$, and the proportion providing shallow-level features, $p_s$, with the
remaining users providing deep features. Both $p_m$ and $p_s$ range from 0 to 1 in steps of
0.25, and for each ratio we average over every possible assignment of depth levels across
the four users. We compare (1) our proposed SIM-CI, (2) two single-modality baselines (Image-only, Text-only), (3) three standard multimodal GNN baselines GCN \cite{kipf2016semi}, GAT \cite{velivckovic2017graph} and GraphSAGE \cite{hamilton2017inductive}
, which replace the proposed structural-bias attention with their standard
aggregation rules.

Under ideal conditions, when all users provide deep features, our model achieves the highest mAP at 0.723. As $p_m$ and $p_s$ increase, every baseline degrades, and the degradation is markedly sharper once shallow features dominate, since shallow features retain mostly local texture and lexical information with a
wide representational gap to deep concepts. In contrast, the SIM-CI remains the most robust across both settings, and its advantage over the baselines widens as the shallow-feature population grows, reaching 0.598 mAP at $p_s=1$ while the strongest
baseline, GAT, falls to 0.537. The GCN, GAT, and GraphSAGE baselines do not explicitly account for feature extraction depth, which limits their ability to handle the semantic differences between shallow and deep features.

This experiment confirms that the tri-heterogeneity adapter is key to solving model-depth
heterogeneity in its most realistic form. The quality-aware contrastive objective guides the adapter to align
features from different extraction depths, reducing the performance
degradation as the proportion of shallow-feature users increases.

\subsection{Performance under Modality Heterogeneity}

To further evaluate the generalizability of SIM-CI beyond the image-text setting used for illustration in Section IV-B, we introduce speech from SpokenCOCO as an additional modality. For each speech sample, we first apply the Montreal Forced Aligner (MFA) \cite{mcauliffe2017montreal} to determine word-level time boundaries. Within each aligned segment, a frozen pre-trained HuBERT-base model \cite{hsu2021hubert} extracts a 768-dimensional acoustic feature vector. The image, text, and speech features are then projected into dimensional of $d=1024$ before being processed by the proposed adapter. We evaluate six client
configurations, Image-only, Text-only, Speech-only, Image+Text, Image+Text+Speech, and a
mixed client population in which each of the $K$ users independently draws one of the five
preceding configurations. For a client with only a subset of modalities, the adapter
retains only the nodes and relations for the available modalities, and multi-modal clients
form cross-modal connections between every available modality pair. All configurations
share the same beamforming network, OTA aggregation, and SIM receiver. We compare SIM-CI
against Mean Pool, which averages all valid tokens directly, and three standard graph
baselines, GCN, GAT, and GraphSAGE, which reuse the same graph nodes and input projections
and only replace the aggregation rule.

Table \ref{tab:client_modality_map} reports mAP for every configuration. Our proposed
SIM-CI achieves the highest mAP in every configuration, and its advantage over the
baselines is most pronounced once more than one modality is available. Adding modalities
consistently improves performance, Image+Text reaches 0.714 mAP and Image+Text+Speech
reaches the highest overall mAP of 0.718, while the mixed client population reaches 0.683,
below the full tri-modal configuration but above every single-modality configuration.
Speech-only gives the lowest mAP among the single-modality configurations, reflecting the
acoustic variability and temporal alignment error in speech representations and their
larger semantic distance to the visual class labels.

Mean Pool ignores the structural relations between tokens and is the weakest method in
every configuration. GCN, GAT, and GraphSAGE recover part of this gap through graph
propagation, but their aggregation mechanisms still cannot fully capture the cross-modal
relations, image-text, image-speech, and text-speech, that the proposed adapter models
explicitly. This is why our advantage over the best baseline grows from single-modality to
multi-modal and mixed configurations, confirming that the tri-heterogeneity adapter, not
just the availability of more data, is what maps heterogeneous client types into a common
semantic space.

\begin{table}[t]
\centering
\caption{mAP under different client modality configurations.}
\label{tab:client_modality_map}
\resizebox{1\columnwidth}{!}{
\begin{tabular}{lccccc}
\toprule
\textbf{Client Configuration}
& \textbf{Mean Pool}
& \textbf{GCN}
& \textbf{GAT}
& \textbf{GraphSAGE}
& \textbf{SIM-CI} \\
\midrule
Image-only
& 0.594 & 0.634 & 0.647 & 0.640 & \textbf{0.655} \\

Text-only
& 0.566 & 0.582 & 0.595 & 0.578 & \textbf{0.602} \\

Speech-only
& 0.502 & 0.521 & 0.534 & 0.527 & \textbf{0.543} \\

Image+Text
& 0.662 & 0.682 & 0.697 & 0.679 & \textbf{0.714} \\

Image+Text+Speech
& 0.681 & 0.694 & 0.701 & 0.688 & \textbf{0.718} \\

Mixed client population
& 0.629 & 0.660 & 0.669 & 0.657 & \textbf{0.683} \\
\bottomrule
\end{tabular}
}
\end{table}
\section{Conclusion} \label{Sec:conclusion}
In this paper, we propose SIM-CI, a novel framework that deploys SIM to perform efficient CI by embedding computation directly into the physical wireless channel. By aggregating multi-user OTA features in a single wave propagation, augmented with a channel-aware robustness adapter and a beamforming network, SIM-CI effectively addresses data, model, and system heterogeneity. Experimental results show that the proposed method achieves accuracy comparable to digital-domain baselines, reaching 98.6\% of the accuracy of conventional digital-domain CI approaches, while reducing
computational energy by 25\% and requiring 75\% fewer transmission time slots, thereby presenting a promising pathway for scalable and low-latency edge intelligence in 6G systems.

\bibliographystyle{ieeetr}
\bibliography{ref}

\end{document}